\documentclass[A4paper,11pt]{article}
\usepackage{graphicx}
\usepackage{psfrag}

\begin{document}

\begin{center}
{\large \bf LARES/WEBER-SAT, frame-dragging and fundamental physics}\\
\end{center}
\vspace{28pt}

\begin{centering}
I. Ciufolini\\
\end{centering}
\vspace{.25in}

Dipartimento di Ingegneria dell'Innovazione, Universit\`{a} di
Lecce, Via Monteroni, \indent 73100 Lecce, Italy\\


{\bf Abstract}

After a brief introduction on the scientific objectives of the LARES/WEBER-SAT
satellite we present the recent measurement of the Lense-Thirring
effect using the nodes of the LAGEOS and LAGEOS 2 satellites and using
the Earth gravity model EIGENGRACE02S obtained by the GRACE space mission, we also include some determination of the rate of change of the lowest order Earth's even zonal harmonics. Finally, we describe an interesting possibility of testing the Brane-World unified theory
of fundamental interactions by the use of a specially designed LARES/WEBER-SAT
satellite.

\section{Introduction}

The scientific objectives of the WEBER-SAT satellite
are:

(1) High precision tests of Einstein's theory of
general relativity, in particular, (1a) a $\sim 1  \%$ measurement of the frame-dragging effect
due to the angular momentum of a body, i.e., the Lense-Thirring
effect, and test of the Earth's gravitomagnetic field. The
Lense-Thirring effect \cite{lent,thi} is a tiny shift of the orbit of a
test-particle.  Frame-dragging, gravitomagnetic field and
Lense-Thirring effect are theoretical predictions of Einstein's
theory of general relativity (see, e.g., ref. \cite{ciuw}).
(1b) A High precision test \cite{ciu04} of large distance infrared
modification of gravity recently proposed by Dvali \cite{dva} to explain the
dark-energy problem (a study will be required to precisely
assess the achievable level of accuracy), see below. An improved
high precision test of the inverse square law for very weak-field
gravity and improved test of the equivalence principle (see the
Italian Space Agency LARES phase A study \cite{ciup}). A $\sim 10^{-3}$ measurement of the
gravito-electric general relativistic perigee precession of the
WEBER-SAT and a high precision measurement of the corresponding
combination of the PPN (Parametrized-Post-Newtonian) parameters
$\beta$ and $\gamma$  \cite{mist,wei}. This test will be achieved in the field of
Earth with a range of about 10000 km, more accurate tests of
$\beta$ and $\gamma$ are achieved in the field of Sun but with a
much longer range. The PPN parameters $\beta$ and $\gamma$ test
Einstein's theory of gravitation versus other metric theories of
gravitation. (1d) Other tests of general relativity and gravitation
(see LARES phase A study \cite{ciup}).

(2) Measurements and improved
determinations in geodesy and geodynamics\\

Frame-dragging may be thought of as an aspect of the Einstein's principle of equivalence
stating that, in a sufficiently small neighbourhood of a spacetime
point, the effects of gravitation are not observable inside a
freely falling frame, i.e. inside the so-called ''Einstein
elevator"  \cite{ciuw}. The basic aspect of the equivalence principle is the
equality of inertial and gravitational mass, which is one of the
best tested principles of physics, measured to date with an
accuracy of about $10^{-13}$. However, the axes of a freely
falling frame, where the equivalence principle holds, are not
fixed relative to ''distant inertial space", i.e. with respect to
''distant fixed stars", but they are ''dragged" by any moving
mass. For example, they are dragged by a rotating mass; this is
the ''dragging of inertial frames" or ''frame-dragging" as
Einstein called it in 1913. In the near future the Gravity Probe B
(GP-B) mission will try to measure frame-dragging, with
unprecedent accuracy, on small super-conducting gyroscopes (the
axes of the frames where the equivalence principle holds) orbiting
around the Earth. GP-B will collect data, over a period of about
one year only, that will then be analyzed to measure
frame-dragging. However, the WEBER-SAT will collect data for a
period of virtually hundreds of years (being a totally passive
satellite with a very small orbital decay). These data could then
be analyzed again in the future using the future improved
gravitational models, in order to obtain much improved tests of
frame-dragging and of other gravitational effects.

 The orbit of a test-particle, such as a small satellite, is
also a kind of gyroscope. Indeed, two of the orbital elements of a
test-particle behave as ''gyroscopes": the node and the pericenter
(neglecting all the other perturbations).

Frame-dragging, gravitomagnetic field and Lense-Thirring effect
have been described in several papers and studies, see for example
ref. \cite{ciuw} and the ASI LARES phase A study  \cite{ciup}.

Here we just point out that the Gravity Probe B experiment will
try to measure the gravitomagnetic effect generated by the angular
momentum of the Earth on a {\it gyroscope}, whereas the WEBER-SAT
should measure the Earth's angular momentum effect on the orbit of
a test particle. In some alternative theories the two effects
may be different as in the case of a non-metric theory with
asymmetric connection, such as the Cartan theory with torsion,
that may affect in a different way the orbit of a test-particle
and of a gyroscope (see ref.  \cite{ciuw}).

\section{A recent measurement of the Lense-Thirring effect
using the LAGEOS satellites}

Let us briefly report a recent measurement of the Lense-Thirring effect
on the two Earth satellites LAGEOS and LAGEOS 2 \cite{ciupav}. We
measured the Earth frame-dragging to be  99 \% of the value
predicted by general relativity; the uncertainty of this
measurement was $± 5$ \% including all the known errors and $± 10$ \%
allowing for underestimated and unknown error sources.

Recently, by analysing the uncertainties in the spherical harmonic
coefficients of the recent Earth gravity model EIGEN-GRACE02S
obtained by the NASA space mission GRACE \cite{rei,tap}, we
found that the only relevant uncertainty in the orbit of the
LAGEOS satellites \cite{ben}, comparing it with the magnitude of the
Lense-Thirring effect, is the one, $\delta J_2$, in the Earth
quadrupole moment, $J_2$, which describes the Earth oblateness. In
the EIGEN-GRACE02S model, the relative uncertainty $\delta
J_2/J_2$ is about $10^{-7}$. This uncertainty corresponds on the
orbits of the LAGEOS satellites to a  shift of the node larger
than a few times the Lense-Thirring effect. However, the orbital
uncertainty due to all the other harmonics is only a few percent
of the general relativity shift. Therefore, in order to eliminate
the orbital uncertainty due to $\delta J_2$ and in order to solve
for the Lense-Thirring effect, it is necessary and sufficient to
use only two observables. The two orbital observables we have
analyzed are the two nodes of the LAGEOS satellites \cite{ciu86,ciu89,ciu96}. After
modelling all the orbital perturbations, apart from the
Lense-Thirring effect, we are able to predict the LAGEOS
satellites' orbit with an error (root-mean-square of the
residuals) of about 3 cm for a 15 day arc, corresponding to about
fraction of  a half millisecond of arc at the LAGEOS satellites
altitude. The Lense-Thirring effect is in contrast 31
milliarcsec/yr on the LAGEOS node and 31.5 milliarcsec/yr on the
LAGEOS 2 node, as calculated by the Lense-Thirring formula. The
residual (calculated minus observed) nodal rate of the LAGEOS
satellites, $\dot{\Omega}_{residual}$ , is therefore: (residual nodal rate) = (nodal rate
from $\delta J_2$ error) + (nodal rate from other $\delta J_{2n}$
errors) + (Lense-Thirring effect) + (other smaller modelling
errors), where the $\delta J_{2n}$ are the errors in the Earth
even zonal harmonic coefficients, $J_{2n}$, of degree $2n$. We can
then solve for the Lense-Thirring effect the system of the two
observed residual nodal rates for the Lense-Thirring effect and
simultaneously eliminate the error due to the $\delta J_{2n}$  uncertainty. The
maximum error in the combination of the residuals due to the
$\delta J_{2n}$ is 4 $\%$ of the Lense-Thirring effect.

         In ref. \cite{ciupav} is reported the analysis (using the orbital
estimator GEODYN \cite{pavd}) of nearly eleven years of laser-ranging data,
         from January 1993 to December 2003, corresponding to about
         one million of normal points, i.e., to about 100 million
         laser ranging observations from more than 50 ILRS stations
         distributed all over the world \cite{ILRS}.

In \cite{ciupav} we analyze the observed residuals of the nodal
longitudes of the LAGEOS satellites, combined according to the
formula to cancel the $\delta J_{2}$ uncertainty. The best fit
line through the raw residuals (one-parameter fit) has a slope of
47.4 milliarcsec/yr; the root-mean-square of these post-fit
residuals is 15 milliarcsec. The residuals after removal of six
main frequencies, corresponding to a thirteen-parameter fit with a
secular trend plus phase and amplitude of six main signals with
periods of 1044, 905, 281, 569, 111 and 284.5 days, have a secular
trend is 47.9 milliarcsec/yr, however the root-mean-square of
these post-fit residuals is 6 milliarcsec only. The Lense-Thirring
effect predicted by general relativity for the combination of the
LAGEOS nodal longitudes amounts to 48.2 milliarcsec/yr. Therefore,
corresponding to the thirteen parameter fit, the observed
Lense-Thirring effect is 47.9 milliarcsec/yr, corresponding to 99
\% of the general relativistic prediction. In conclusion, this
analysis confirms the Einstein's theory predictions of
frame-dragging and Lense-Thirring effect \cite{ciupav}. The total
uncertainty
         of our measurement is, including systematic errors,
         $± 5 \%$ of the Lense-Thirring effect and $± 10 \%$ allowing
         for underestimated and unknown error sources. For example, if
         we consider the time-independent gravitational error
         (root-sum-square) to be three times larger we get a
         corresponding error of 9 \% and a total uncertainty
         of less than 10 \%.

\section{Some geodetic results using the LAGEOS satellites and
the EIGENGRACE02S model}

Using the method of analysis of about 11 years of satellite ranging observations reported in the previous section,
in addition to the accurate determination of the Lense-Thirring effect, an anomalous variation in the Earth gravity field since 1998 was observed \cite{ciupavper} that was clearly identified as an anomalous increase in the Earth quadrupole moment. The trend in the nodal longitudes of both satellites distinctly showed a variation in the Earth gravity field since 1998. This effect was proved \cite{ciupavper} to be due to an increase in the $J_2$  coefficient, indeed, combining the node residuals according to the formula to eliminate the $J_2$  perturbation only, the effect disappeared. The anomalous trend observed using EIGEN-GRACE02S was also accurately reproduced using the previous EGM96 Earth gravity model and the recent EIGEN-2 model due to the CHAMP satellite. This result confirms the measured relative increase of  $J_2$ of the order of $10^{-11}$ that was recently reported \cite{coxc}. The Earth mass redistribution associated with this phenomenon is so far not clearly understood.

It is important to stress that together with the measurement of
the Lense-Thirring effect, it was also measured the effect of the
variations of $J_2$, $J_4$ and $J_6$ on the nodes of the LAGEOS
satellites. In \cite{ciu2005} it is indeed reported an effective
(i.e. including the effect of the higher even zonal harmonics)
value of $\dot J_4^{Effective} \cong - 1.5*10^{-11}$. In the
EIGENGRACE02S model, obtained by the GRACE mission only, the Earth
gravity field was measured during the period 2002-2003.
Corrections due to $\dot J_2$ and $\dot J_4$ were then applied to
this 2002-2003 measurement in order to obtain a gravity field
model antecedent to 2002-2003. These values of $\dot J_2$ and
$\dot J_4$ used by the GFZ team are $\dot J_2 = - 2.6*10^{-11}$ and
$\dot J_4 = - 1.41*10^{-11}$ and they were measured on the basis of
completely independent 30-year observations before 2002.
The only constraint (based on the validity of the GRACE
measurements) is that the $\dot J_4$ correction applied to $J_4$,
{\it must} of course produce in 2002-2003 the same value of $J_4$
that was measured by GRACE in 2002-2003,
at least within the EIGENGRACE02S uncertainties in $J_{4}$.
In ref. \cite{ciu2005} are reported the orbital analyses using the orbital
estimator GEODYN with and without a contribution of $\dot J_4 =
- 1.41*10^{-11}$. First, it is important to stress that in the case
of {\it not} applying this $\dot J_4$ correction to the orbital
analysis, it can be clearly seen, by visual inspection, a hump in
the combined residuals. Indeed, the effect of the time variation
$\dot J_4$ shows up as a quadratic effect in the cumulative nodal
longitude of the LAGEOS satellites, therefore the combined
residuals of LAGEOS and LAGEOS 2 were fitted with a parabola,
together with a straight line and with the main periodic terms.
Then, by fitting the raw residuals obtained {\it without} any
$\dot J_4$, it was measured a $\dot J_4^{Effective} \cong
- 1.5*10^{-11}$, which includes the effect of $\dot J_6$ and of
higher even zonal harmonics on the LAGEOS satellites. On other hand, in the analysis of the
combined residuals obtained with the EIGENGRACE02S correction of
$\dot{J_4}= - 1.41*10^{-11}$, it was measured a $\dot
J_4^{Effective}$ of less than $ - 0.1*10^{-11}$, in complete
agreement with the previous case. It is finally important to
stress that this small value of the unmodelled quadratic effects
in our nodal combination due to the unmodelled $\dot J_{2n}$
effects (with $2n \geq 4$) corresponds to a change in the measured
value of the Lense-Thirring effect of about 1 $\%$. In other words
using the value of $\dot J_4^{Effective} = - 1.5*10^{-11}$ for the LAGEOS satellites that we
obtained from fitting the combined residuals (which is about 6
$\%$ larger than the value $\dot J_4 = - 1.41*10^{-11}$ given in the
EIGENGRACE02S model) resulted in a change of the measured value of
frame-dragging by about 1 $\%$ only with respect to the case of
using $\dot J_4 = - 1.41*10^{-11}$; in conclusion this 1 $\%$
variation fully agrees with the error analysis given in
\cite{ciupav}.

\section{On the possibility of testing Brane-World theories with
WEBER-SAT/LARES}

Let us now briefly describe the possibility of probing some
recently proposed modifications of gravity using the Runge-Lenz
vector, i.e., the perigee of WEBER-SAT \cite{ciu04}.

In Newtonian mechanics, the orbital angular momentum of a
satellite and its nodal line, the intersection of its orbital
plane with the equatorial plane of the central body, maintain a
constant direction relative to ''distant inertial space" for a
motion under a central force. The Runge-Lenz vector, joining the
focus and the pericenter of the orbit of a satellite, has also a
constant direction relative to "distant inertial space" for a
motion under a central force dependent on the inverse of the
squared distance from the central body. Using the technique of
laser-ranging with retro-reflectors to send back the short laser
pulses, to this date we are able to measure distances with a precision
of a few cm to a point on the Moon and of a few
millimeters to a small artificial satellite. The instantaneous
position of the LAGEOS satellites can be measured with an
uncertainty of a few millimeters and their orbits, with semi-major
axes $a_{LAGEOS} \cong 12270$ km and $a_{LAGEOS \; II} \cong
12210$ km, can be predicted, over 15 day periods, with a
root-mean-square of the range residuals of a few cm. This
uncertainty in the calculated orbits of the LAGEOS satellites is
due to errors in modelling their orbital perturbations and, in
particular, in modelling the deviations from spherical symmetry of
the Earth's gravity field, described by a spherical harmonics
expansion of the Earth's potential. However, to date, the
terrestrial gravity field is determined with impressive accuracy,
in particular with the dedicated satellites CHAMP and especially
GRACE \cite{rei}. Regarding the perigee, the observable quantity is $e a \dot
\omega$, where $e$ is the orbital eccentricity of the satellite
and, $\omega$, the argument of perigee, that is the angle on its
orbital plane measuring the departure of the satellite perigee
from the equatorial plane of Earth. Therefore, we can increase the
measurement precision by considering orbits with larger
eccentricities.

Motivated by the cosmological dark energy problem, Dvali recently
proposed string theories leading, among other things, to weak
field modifications of gravity \cite{dva}. One of the interesting
observational consequences of the large distance infrared
modification of gravity pointed out by Dvali is the anomalous
shift of the pericenter of a test particle. The anomalous
perihelion precession predicted by this gravity modification for
the Moon perigee is:

\begin{equation}
\delta \phi = - [(3 \pi \sqrt{2} /4) r^{3/2}]/({r_c} {r_g}^{1/2})
\; rad/orbit \end{equation}

Where, $r_g = 0.886cm$ is the gravitational radius of the Earth,
$r$ is the Earth-satellite distance and $r_c = 6 Gpc$ is the
gravity modification parameter that gives the observed galaxies
acceleration without dark energy \cite{dva}. $ \delta \phi =
1.4 \cdot 10^{-12}$ rad/orbit for the Moon.

Therefore in the case of the WEBER-SAT satellite with a semimajor
axis of about 12270 km this effect would amount to 0.04
milliarcsec/yr only.

Since this effect of infrared gravity modification is proportional
to the 3/2 power of the semi-major axis and however the number of
orbits per year goes as the -3/2 power of the semi-major axis, in
terms of radians per year the perigee shift is the same for both
the Moon and WEBER-SAT, i.e. $1.9 \cdot 10^{-11}$ rad/yr.
Therefore, we \cite{ciu04} simply need to consider what can be gained, or lost,
with the use of WEBER-SAT versus the Moon. In regard to the measurement precision,
the ranging precision  is very roughly
proportional to the range distance, i.e. is a few cm for the Moon
and a few millimeters for the WEBER-SAT, then since the shift of
the perigee at the satellite altitude is $1.9 \cdot 10^{-11}$
rad/yr times the semi-major axis, the ratio of ranging precision to
the effect to be measured is roughly the same for both the Moon
and WEBER-SAT, even though slightly more favorable for the Moon.
However, since the recovery of the perigee shift is proportional
to the eccentricity of the satellite, we could orbit the WEBER-SAT
satellite with a much larger eccentricity than the one of the Moon
and therefore we could make the measurement of the perigee shift
of the WEBER-SAT more {\it precise} than the one of the Moon.

However, critical are the {\it systematic errors} acting on the
WEBER-SAT:

(a) the impact of the modelling uncertainties in the gravitational
perturbations is critical for the WEBER-SAT satellite indeed the
zonal harmonics of the Earth gravity field produce a perigee shift
that is a function of the inverse powers of the semi-major axis,
$a$: $1/{a^{(2n+3/2)}}$, for each even zonal harmonic coefficient
$J_{2n}$ with $n$ integer. However, considering the improvements
in the future Earth gravity models from the mission CHAMP, GRACE
and GOCE, the future uncertainty in the perigee shift due to
gravitational perturbations should drastically decrease.

There is also an interesting possibility to choose a different
orbit for WEBER-SAT, not at 70 degrees of inclination, but for
example with the special inclination of about 63.4 degrees, (of
the type of the Molniya orbits) at which a satellite would have a
null perigee shift due to the Earth's quadrupole moment $J_2$. In
this way we would be able to cancel the major part of the $J_{2n}$
uncertainties to measure the perigee, but we will lose some
accuracy in the measurement of the Lense-Thirring effect on the
node, even though the use of LAGEOS and LAGEOS II, together with
WEBER-SAT, and the future improvements in the accuracy of the
Earth's gravity models should not make this Lense-Thirring
measurement much worst than a 1 $\%$ measurement (this possibility
has to be further investigated) and furthermore, in this case, the Lense-Thirring effect could be
measured using the WEBER-SAT perigee.

The impact of the modelling uncertainties in the non-gravitational
perturbations is also critical for the WEBER-SAT satellite indeed
the crucial factor is the cross-sectional area to mass ratio. In
other words the acceleration produced by a non-gravitational force
(such as radiation pressure) acting on a satellite is proportional
to its cross sectional area and inversely proportional to its
mass. Then this ratio is roughly proportional to the inverse of
the radius of the considered satellites. Then for the Moon is very
small whereas for WEBER-SAT may be a critical source of error.

These forces could be reduced by (a) a much denser and larger
satellite than the one of the original proposal (although
by increasing the cost of the mission); (b) the non-gravitational
perturbations of the perigee could also be reduced through the use of
a much more eccentric orbit and (c) the mismodelling of the
radiation pressure perturbations could be reduced by special
optical and thermal tests performed on the WEBER-SAT satellite and
through the measurement of its spin axis and rotational rate, once
in orbit. Finally, we also mention the possibility of a drag free
system, that implies acceleration sensors and propulsion systems on
board, or of a spherical satellite made of a material transparent to
most of the radiation; similar spherical retro-reflectors have
been tested by the Russian space agency.

On September 2004 at a LARES INFN-meeting in Padova (with B.
Bertotti, M. Cerdonio, F. Morselli, A. Riotto, I.C., A. Paolozzi,
S. Dell'Agnello, L. Iorio and others.), I.C. pointed out the need
to use a very large eccentricity and a higher orbit of LARES in
the attempt to measure the Brane-World effects on its orbit. In
this way, with a larger value of the eccentricity, $\it e$, the
non-gravitational effects on its perigee would be reduced.
Contemporarily, using a higher orbit all the effects due to
$J_{2n}$ with ${2n} \geq 4$ would be drastically reduced. At that
meeting I.C. proposed, {\it as an example}, an orbit with
semimajor axis of 36000 km. A very large value of the eccentricity
is critical to minimize the non-gravitational perturbations on the
perigee, on other hand a large value of the semimajor axis with a
suitable value of $\it e$ would reduce the $\delta J_{2n}$ effects
on the perigee and make the errors due to the $\delta J_{2n}$,
with $2n \geq 4$, smaller than the Brane-World effect on the LARES
perigee.

The suitable orbit for LARES is currently under study, indeed as pointed
out in ref. \cite{ciupcp} a higher orbit would imply the need of a larger satellite
in order for LARES to be visible by the laser ranging stations,
however this will imply a higher cost for the satellite and for its launch.

\end{document}